\documentclass[letterpaper]{article} 
\usepackage{aaai25}  
\usepackage{times}  
\usepackage{helvet}  
\usepackage{courier}  
\usepackage[hyphens]{url}  
\usepackage{graphicx} 
\usepackage{natbib}  
\usepackage{caption} 
\usepackage{lineno}
\usepackage{colortbl}
\usepackage{amsmath,amssymb,amsfonts}
\usepackage{xcolor}
\usepackage{booktabs}  
\usepackage{threeparttable}  
\usepackage{multirow}
\usepackage{footnote} 
\usepackage[most]{tcolorbox}

\definecolor{gray}{RGB}{128,128,128}

\usepackage{algorithm}
\usepackage{algorithmic}

\usepackage{newfloat}
\usepackage{listings}
\DeclareCaptionStyle{ruled}{labelfont=normalfont,labelsep=colon,strut=off} 
\floatstyle{ruled}
\newfloat{listing}{tb}{lst}{}
\floatname{listing}{Listing}
\title{Understanding Emotional Body Expressions via Large Language Models}

\author{
    Haifeng Lu\textsuperscript{\rm 1,2},
    Jiuyi Chen\textsuperscript{\rm 3,4},
    Feng Liang\textsuperscript{\rm 1},
    Mingkui Tan\textsuperscript{\rm 3},
    Runhao Zeng\textsuperscript{\rm 1,5$\ast$},
    Xiping Hu\textsuperscript{\rm 1,6}\thanks{Corresponding author}
}
\affiliations{
    \textsuperscript{\rm 1}Guangdong-Hong Kong-Macao Joint Laboratory for Emotional Intelligence and Pervasive Computing, \\  Shenzhen MSU-BIT University,
    \textsuperscript{\rm 2}Lanzhou University,
    \textsuperscript{\rm 3}South China University of Technology,\\
    \textsuperscript{\rm 4}Peng Cheng Laboratory,
    \textsuperscript{\rm 5}Shenzhen University,
    \textsuperscript{\rm 6}Beijing Institute of Technology\\
    luhf18@lzu.edu.cn, zengrh@smbu.edu.cn, huxp@smbu.edu.cn
}

\usepackage{bibentry}

\begin{document}

\maketitle

\begin{abstract}
	
	Emotion recognition based on body movements is vital in human-computer interaction. However, existing emotion recognition methods predominantly focus on enhancing classification accuracy, often neglecting the provision of textual explanations to justify their classifications. In this paper, we propose an \textbf{E}motion-\textbf{A}ction \textbf{I}nterpreter powered by \textbf{L}arge \textbf{L}anguage \textbf{M}odel (EAI-LLM), which not only recognizes emotions but also generates textual explanations by treating 3D body movement data as unique input tokens within large language models (LLMs). Specifically, we propose a multi-granularity skeleton tokenizer designed for LLMs, which separately extracts spatio-temporal tokens and semantic tokens from the skeleton data. This approach allows LLMs to generate more nuanced classification descriptions while maintaining robust classification performance. Furthermore, we treat the skeleton sequence as a specific language and propose a unified skeleton token module. This module leverages the extensive background knowledge and language processing capabilities of LLMs to address the challenges of joint training on heterogeneous datasets, thereby significantly enhancing recognition accuracy on individual datasets. Experimental results demonstrate that our model achieves recognition accuracy comparable to existing methods. More importantly, with the support of background knowledge from LLMs, our model can generate detailed emotion descriptions based on classification results, even when trained on a limited amount of labeled skeleton data.

\end{abstract}

\section{Introduction}

In the field of human-computer interaction, a machine's ability to understand human emotions directly impacts the user's interaction experience \cite{fragopanagos2005emotion, mandryk2006continuous, narayanan2020proxemo}. Traditionally, the most common methods for emotion recognition involve analyzing facial expressions and vocal tones \cite{Li2022facialSurvey, el2011survey}. However, accurately recognizing a user's emotions becomes particularly challenging when the user is distant from the camera or when a microphone is unavailable.

Numerous studies have shown that full-body motion is a significant means of expressing emotion \cite{noroozi2018survey, wang2023emotion}. Additionally, the large surface area of the human torso facilitates data collection from a distance, laying the foundation for more extensive emotion recognition scenarios \cite{wang2023emotion}. With recent advancements in motion capture technology and human pose estimation algorithms \cite{peng2024dual, zheng2020deep, zhang2012microsoft}, the acquisition of 2D and 3D skeleton data has become more convenient and efficient. Notably, 3D skeleton data offer the advantage of being resistant to variations in viewpoint, lighting, and background clutter \cite{Lu_2023}. As a result, emotion recognition based on 3D full-body skeleton data has garnered significant attention in recent years.

Most existing skeleton-based emotion recognition methods rely on handcrafted features \cite{fourati2019contribution, daoudi2017emotion, ouguz2024emotion}. In addition, deep learning approaches such as those in \cite{Beyan2023TAC}, ST-Gait++ \cite{lima2024st}, and BPM-GCN \cite{10433680} are also significant for learning emotion representations. However, these methods share some common limitations: (i) Traditional emotion recognition techniques primarily focus on improving classification accuracy and often lack the capability to provide textual explanations that support their classifications. (ii) Due to variations in data collection methods, 3D skeleton datasets are often heterogeneous, with differences in the number of joints and frame lengths. This heterogeneity impedes effective knowledge transfer between datasets and restricts the potential for improving classification accuracy within individual datasets.

\textbf{Motivation.} Inspired by the remarkable capabilities of large language models (LLMs) across various domains \cite{li2024mvbench, qu2024llms}, we are intrigued by the potential of utilizing LLMs as emotion recognizers that can not only classify emotions from skeleton data but also generate corresponding classification criteria. Several studies have shown that LLMs hold some features that are useful for emotion recognition. Specifically, these models are pre-trained on vast corpora that include descriptions of emotional actions \cite{qiu2024language, li2023large}. Therefore, converting 3D full-body skeleton data into tokens recognizable by LLMs, and utilizing the extensive background knowledge of LLMs to classify emotions and generate detailed explanations, would be highly valuable.

To address these challenges, we propose an \textbf{E}motion-\textbf{A}ction \textbf{I}nterpreter powered by \textbf{L}arge \textbf{L}anguage \textbf{M}odel (EAI-LLM), which is capable of simultaneously recognizing emotions from skeleton data and generating detailed emotion descriptions. First, we introduce a Multi-Granularity Skeleton Tokenizer (MGST) to enhance token diversity, enabling LLMs to produce more fine-grained emotion descriptions. To further improve performance, we implement a Unified Skeleton Token (UST) module that incorporates a spatio-temporal masking mechanism. This module not only addresses the challenges posed by heterogeneous datasets but also enhances recognition accuracy on individual datasets. Additionally, we pre-train the EAI-LLM using a combination of skeleton-language data and fine-tune it on prompt-based question-and-answer tasks. To convert skeleton features into tokens recognizable by LLMs, we also introduced a skeleton-text contrastive learning framework. This approach aligns the skeleton feature space with the semantic space using contrastive loss. Experimental results demonstrate that the recognition capability of EAI-LLM is comparable to existing methods, and it can generate detailed emotion descriptions from classification results by leveraging the background knowledge of LLMs.

In summary, our main contributions include:

\begin{itemize}
	
	\item We propose a novel 3D full-body skeleton-based emotion recognition model called EAI-LLM. To the best of our knowledge, EAI-LLM is the first approach to utilize LLMs as emotion recognizers while generating detailed emotion descriptions directly from skeleton data.
	
	\item We propose a Multi-Granularity Skeleton Tokenizer (MGST), which can enhance token diversity, with the goal of improving the model's ability to infer emotion expressed through 3D skeleton movement sequences.
	
	\item We propose an Unified Skeleton Token (UST) module designed for joint training in heterogeneous datasets. It treats skeleton sequences as a specialized language, allowing heterogeneous datasets to be unified within a shared language space for training.
	
\end{itemize}

\section{Related Work}\label{releated_work}

\subsection{Body Skeleton-based Emotion Recognition}

Many studies have focused on using body movements, postures, and gestures to recognize emotion. Traditionally, emotion recognition from body movements has relied on handcrafted features \cite{piana2016adaptive, daoudi2017emotion, fourati2019contribution}. For instance, \cite{4563173} examined upper body motions, utilizing features such as the number of local maxima and the ratio between the maximum and the duration of the largest peaks to identify emotions. Similarly, \cite{7163145, fourati2019contribution} proposed a full-body motion encoding scheme that extracted over 110 motion features across three levels, anatomy, spatial direction, and posture/movement, to describe expressive movements. Additionally, \cite{ouguz2024emotion} explored commonly used time, frequency, and statistical-based parameters, employing feature selection methods to select four features from each frame to construct a feature matrix for emotion recognition.

Recently, due to the impressive performance of neural networks \cite{zeng2024improving, you2024toward}, deep learning models have dominated the field of emotion recognition from 3D skeleton movement sequences.  \cite{ghaleb2021skeleton} proposed a classifier network based on Spatial-Temporal Graph Convolutional Networks \cite{yan2018spatial} to recognize emotions from body movements. In \cite{zhang2021emotion}, the authors introduced an attention-based stacked LSTM network to detect the relationship between emotions and movements. \cite{Beyan2023TAC} encoded various time intervals of 3D positional data into RGB images and then used Convolutional Neural Networks (CNNs) for classification. Similarly, \cite{wang2023emotion} combined handcrafted features with CNN-learned image features and employed a linear classifier for emotion prediction. Unlike these previous models, which focus solely on classification, our method not only accurately identifies emotion from skeleton data but also provides interpretable explanations, significantly broadening the model's application scenarios.

\subsection{Multi-modal Large Language Models}

Recently, LLMs have demonstrated impressive capabilities in generation and comprehension. Early models such as BERT \cite{devlin2018bert}, GPT-2 \cite{radford2019language}, and T5 \cite{raffel2020exploring}, which were trained on web-scale text data, laid the groundwork for the popularity of large models. Subsequently, models with greater capacity, more parameters, and extensive training data, such as GPT-3 \cite{brown2020language}, Alpaca \cite{alpaca}, PaLM \cite{chowdhery2023palm}, Vicuna \cite{vicuna2023}, and LLaMA \cite{touvron2023llama}, have been developed. Recent advancements, such as InstructGPT \cite{ouyang2022training}, have focused on aligning LLMs with human instructions and feedback. ChatGPT \cite{ChatGPT}, Anthropic \cite{anthropic2024claude}, and GPT-4 \cite{GPT4} now interact with users and answer a broad range of diverse and complex questions.

With the impressive generalization abilities of LLMs, some studies have extended these models to other modalities. Flamingo \cite{alayrac2022flamingo} and BLIP-2 \cite{li2023blip} aligned pre-trained vision encoders with language models using a cross-attention mechanism, establishing a foundation for subsequent research in multi-modal LLMs. More recently, the capabilities of LLMs to handle various tasks have been further explored. GPT-4 \cite{GPT4} has demonstrated powerful visual understanding and reasoning abilities, while MiniGPT-4 \cite{zhu2023minigpt} and LLaVA \cite{liu2024visual} produced detailed and accurate image descriptions through a series of visual instruction-tuning datasets. VideoChat \cite{li2023videochat} and VideoChatGPT \cite{maaz2023video} utilized ChatGPT to generate video instruction-tuning data, extending support to video modalities. Despite these advancements, there remains a gap in multi-modal LLMs capable of processing skeleton data. In this work, we propose a new framework that enables LLMs to handle skeleton-based body movement data.

\begin{figure*}[tp]
	\centering
	\centerline{\includegraphics[width=6.8in]{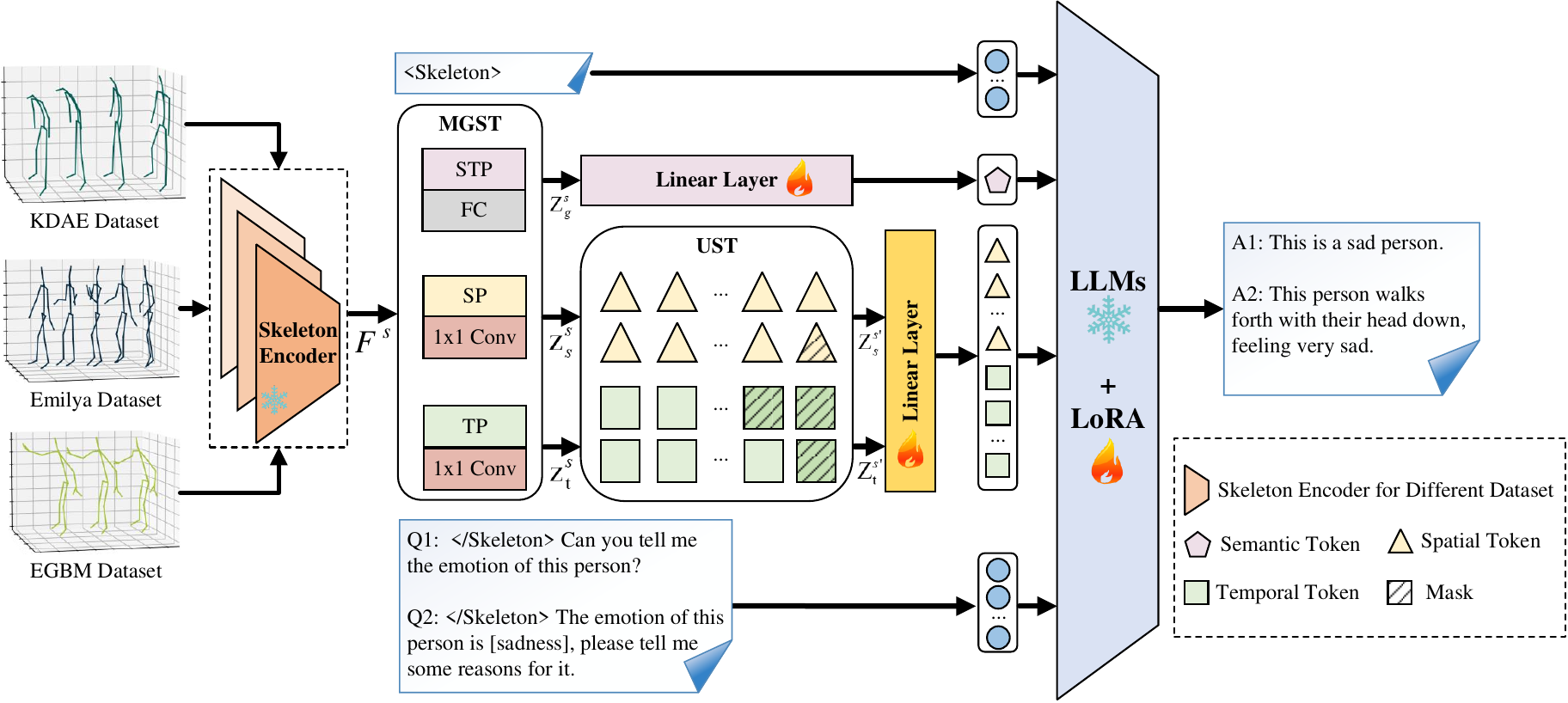}}
	\caption{This work presents a novel approach for 3D full-body skeleton-based emotion recognition using fine-tuned LLMs, termed EAI-LLM. Unlike previous methods, EAI-LLM not only identifies emotions but also generates textual explanations by treating 3D body movement data as unique input tokens within the LLMs.}
	\label{skeleton_figure}
\end{figure*}

\section{Method}\label{Method}

\subsubsection{Problem definition.} Let $X = \{X_{t} \in \mathbb{R}^{T \ast J \ast 3}\}_{t=1}^{T}$ be a 3D skeleton sequence, where $X_{t}$ denotes the frame at time $t$ with $J$ body joint. Our objective is to input a 3D skeleton movement sequence $X$ into the EAI-LLM, which automatically encodes $X$ into semantic space, outputs corresponding emotion labels for the body movement data and provides detailed emotion descriptions.

\subsubsection{Overview.} We propose an emotion recognition framework called EAI-LLM, as shown in Fig. \ref{skeleton_figure}. First, we introduce a multi-granularity skeleton tokenizer that extracts spatio-temporal tokens and semantic tokens distinctly. Second, we employ a masking mechanism to normalize tokens of varying lengths to a unified length, thereby enabling joint training. Third, we use a linear projection layer to bridge the gap between the skeleton encoder and LLMs, embedding the extracted multi-granularity skeleton tokens together with text prompts as input for LLMs. Finally, we pre-train EAI-LLM using a composite of skeleton-language data and fine-tune it on prompt-based question-and-answer tasks.

\subsection{Multi-Granularity Skeleton Tokenizer}\label{MST}

Traditional skeleton-based emotion recognition methods typically use a skeleton encoder to extract features, denoted as $F^{s} \in \mathbb{R}^{T \ast J \ast C}$, where $C$ represents the base channel. These features are then spatio-temporally compressed and passed through a fully connected (FC) layer to obtain classification results. However, the fundamental design of LLMs is to process text-based inputs, making it impossible to directly input skeleton sequence $X$ into LLMs. To bridge this gap, we develop a skeleton tokenizer to convert skeleton data into a format compatible with the input structure of LLMs. 

In this paper, we propose a multi-granularity skeleton tokenizer to enhance token diversity, thereby facilitating the generation of more detailed text descriptions. First, we perform spatio-temporal pooling (STP) on the feature $F^{s}$, followed by processing through a FC layer to obtain a semantic token $z^{s}_{g} \in \mathbb{R}^{C}$. This token encapsulates the global body motion encoding information. While the semantic token is advantageous for classification tasks, the spatio-temporal pooling process can result in information loss, which is detrimental in generation tasks.

To address this, we extract spatio-temporal tokens to enhance token diversity. Specifically, we employ a temporal pooling (TP) layer and a 1$\times$1 convolution layer to aggregate the skeleton features $F^{s}$ across time. Similarly, we apply a spatial pooling (SP) layer and a 1$\times$1 convolution to average the spatial dynamics of all joints. The resulting spatio-temporal features are then flattened separately into unified representations, referred to as the spatial tokens ${z}^{s}_{s} \in \mathbb{R}^{J \ast C}$ and temporal tokens ${z}^{s}_{t} \in \mathbb{R}^{T \ast C}$, respectively.

\subsection{Unified Skeleton Token Module}

Due to variations in data collection methods across different datasets, the number of joints $J$ and frame lengths $T$ can differ significantly among datasets. Consequently, the dimensions of the spatio-temporal tokens extracted in Sec. \ref{MST} are inconsistent. Traditionally, classification tasks using skeleton data have addressed this heterogeneity by training separate models for each dataset. In this paper, we approach skeleton sequences as a specialized form of language, where differences in frame length and joint count are analogous to sentences of varying lengths. To unify these differences, we leverage the masking mechanism from natural language processing by applying a mask to all skeleton tokens, as shown in Eq. (\ref{mask}). This approach allows us to merge all datasets into a larger, unified dataset in the language space for emotion representation learning.
\begin{equation}\label{mask}
\begin{aligned}
{{z}^{s \prime}_{s} } & ={z}^{s}_{s} \odot M^{s}_{s} \\
{{z}^{s \prime}_{t} } & ={z}^{s}_{t} \odot M^{s}_{t},
\end{aligned}
\end{equation}
where $M^{s}_{s}$ and $M^{s}_{t}$ are a $1 \times L$ attention map that serves to retain the original tokens and nullify the padded elements. $L$ represents the maximum length of all tokens. $\odot$ denotes element-wise multiplication.

\subsection{Skeleton-Aware Large Language Model}

LLMs typically take sentences in human language as input instructions, but the skeleton sequences we use are not well-compatible with LLMs. To enable LLMs to recognize skeleton sequences and generate text descriptions, we apply Low-Rank Adaptation (LoRA) \cite{hu2021lora} to LLMs so that it better understands skeleton sequences while keeping the model's pre-trained weights unchanged. Specifically, for each training sample composed of a skeleton sequence and its corresponding emotion label and emotion description, we perform the following steps:
\begin{enumerate}
	
	\item We utilize a pre-trained GCN-based skeleton encoder to obtain the initial skeleton tokens ($z^{s}_{g}$, ${z}^{s\prime}_{s}$, ${z}^{s\prime}_{t}$).
	\item Prior work on multimodal LLMs \cite{liu2024visual} show that a learnable linear layer maps the visual feature space to the linguistic feature space, enabling LLMs to handle non-text data. Following this approach, we use a linear layer to transform the initial 768-dimensional skeleton token into 4096-dimensional token, effectively bridging the gap between the skeleton encoder and LLMs.
    
	\item We design two prompts that conform to the LLaMA-2 conversation template, as shown below. 
	
	\subsubsection{Emotion Recognition}
	\#Human: \textless Skeleton\textgreater{} \textless SkeletonFeature\textgreater{} \textless \slash{}Skeleton\textgreater{} Can you tell me the emotion of this person? \#Assistant:
	
	\subsubsection{Emotion Description}
	\#Human: \textless{}Skeleton\textgreater{} \textless{}SkeletonFeature\textgreater{} \textless{}\slash{}Skeleton\textgreater{} The emotion of this person is [shame], please tell me some reasons for it. \#Assistant:
	
	In these prompts, \textless SkeletonFeature \textgreater represents a skeleton token derived in step 2, and the word within square brackets represents an emotion label.
	\item We concatenate the skeleton tokens with the prompt tokens and input them together into LLMs.
	\item We use Eq. (\ref{LoRA}) to constrain the similarity between the ground truth tokens $t_g$ and the predicted tokens $t_p$, adjusting EAI-LLM through LoRA. 
	\begin{equation}\label{LoRA}
	\mathcal{L}_{LoRA}=\mathcal{L}_{ce}(t_p, t_g),
	\end{equation}
	where $\mathcal{L}_{ce}(.,.)$ is the cross-entropy loss.
	
\end{enumerate}

\subsection{Skeleton-Language Alignment}\label{SLP}

Skeleton and text data exist in distinct feature spaces, making it challenging for LLMs to interpret unaligned skeleton features. To address this, we employ contrastive learning to align the skeleton feature space with the linguistic feature space, thus improving the compatibility of the skeleton tokens with LLMs. The subsequent sections will provide a detailed explanation of the alignment process.

\subsubsection{Skeleton Encoder} Our skeleton encoder is designed in a manner similar to CTR-GCN \cite{chen2021channel}, enabling the extraction of skeleton features $F^{s}$. Following pooling and dimensional transformation, we derive a C-dimensional feature vector $z^{s}$. The backbone of the encoder is flexible and can be substituted with other GCN-based networks, such as HD-GCN \cite{Lee_2023_ICCV} or 2s-AGCN \cite{2sagcn2019cvpr}. 

\subsubsection{Text Encoder} We employ a pre-trained CLIP model (specifically ViT-L/14) \cite{radford2021learning} as the text encoder to extract a feature vector $z^{t}$ representing the text description. Drawing inspiration from ActionCLIP \cite{wang2023actionclip}, we manually design the text descriptions in the format: "\textit{This is a [happy] person.}", where the word within the brackets changes according to the skeleton label.

\begin{figure}[tp]
	\centering
	\centerline{\includegraphics[width=3.2in]{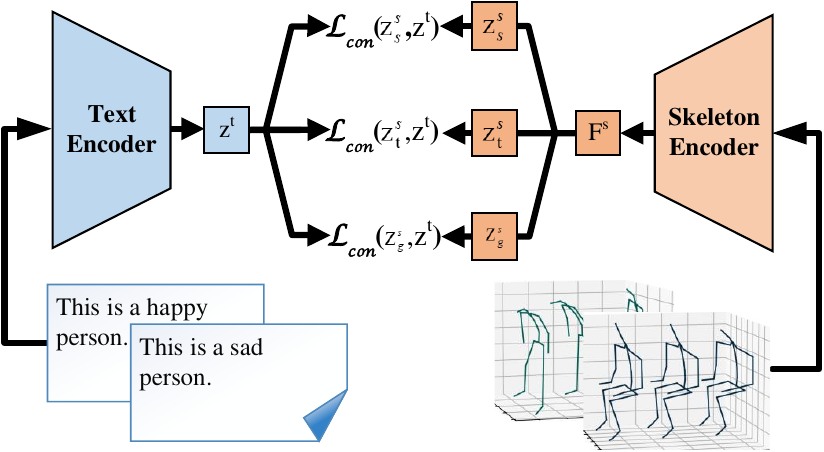}}
	\caption{Diagram of skeleton-language alignment.}
	\label{little_Frame}
\end{figure}

\subsubsection{Semantic Alignment} Contrastive learning excels at unifying data from different modalities into a shared feature space, facilitating cross-modal knowledge transfer \cite{radford2021learning, xiang2023generative}. Consequently, we employ contrastive learning to align the skeleton feature space with the linguistic feature space, enabling the compatibility of these tokens with LLMs.

To bring the pairwise skeleton representation $z^{s}$ and text description label representation $z^{t}$ closer together, we calculate the skeleton-to-text and text-to-skeleton similarity scores as specified in Eq. (\ref{infoNCE}).
\begin{equation}\label{infoNCE}
\begin{aligned}
p_i^{s2t}\left(z^{s}_{i}\right) & =\frac{\exp \left(\operatorname{cos}\left(z^{s}_{i}, z^{t}_{i}\right) / \tau\right)}{\sum_{j=1}^N \exp \left(\operatorname{cos}\left(z^{s}_{i}, z^{t}_{j}\right) / \tau\right)} \\
p_i^{t2s}\left(z^{t}_{i}\right) & =\frac{\exp \left(\operatorname{cos}\left(z^{t}_{i}, z^{s}_{i}\right) / \tau\right)}{\sum_{j=1}^N \exp \left(\operatorname{cos}\left(z^{t}_{i}, z^{s}_{j}\right) / \tau\right)},
\end{aligned}
\end{equation}
where $\operatorname{cos}$ represents cosine similarity, $\tau$ is the temperature hyper-parameter, $N$ represents the batch size. Since the number of skeleton sequences greatly exceeds the number of emotion labels, there will be multiple skeleton sequences with the same label within the same batch. Consequently, using cross-entropy to calculate the similarity between $p_i^{s2t}$ and $p_i^{t2s}$ is not suitable. Therefore, we redefine the skeleton-text contrastive loss using the Kullback–Leibler (KL) divergence, as shown in Eq. (\ref{KL}).
\begin{equation}\label{KL}
\begin{aligned}
\mathcal{L}_{con}\left(z^{s}, z^{t}\right)=\frac{1}{2} \mathbb{E}_{(z^{s}, z^{t}) \sim \mathcal{D}} & \left[\operatorname {KL} \left(p^{s2t}(z^{\mathbf{s}}),\right.\right.  \left.\hat{y}\right) \\
& \left.+\mathrm{KL}\left(p^{t2s}(z^{t}), \hat{y}\right)\right],
\end{aligned}
\end{equation}
where $\mathcal{D}$ is the entire dataset. The ground-truth $\hat{y}$ is defined as 1 for positive pairs and 0 for negative pairs. 

In this paper, we optimize the semantic tokens and spatio-temporal tokens using skeleton-language alignment, making them more easily recognized by LLMs. The loss functions can be represented by Eq. (\ref{ST_token}) and Eq. (\ref{SA_token}), respectively.
\begin{equation}\label{ST_token}
\mathcal{L}_{st}  = \mathcal{L}_{ce}({z}^{s}_{g}, y) + \frac{1}{2}(\mathcal{L}_{con}({z}^{s}_{s}, z^{t}) + \mathcal{L}_{con}({z}^{s}_{t}, z^{t})),
\end{equation}
\begin{equation}\label{SA_token}
\mathcal{L}_{se}  = \mathcal{L}_{ce}({z}^{s}_{g}, y) + \mathcal{L}_{con}({z}^{s}_{g}, z^{t}),
\end{equation}
where \textit{y} is the one-hot presentation of the emotion label.  

\section{Experiment}
\subsection{Datasets}

\subsubsection{Emilya} The EMotional body expression In daILY Actions dataset \cite{fourati2016perception} comprises 8,206 samples. Eleven actors performed eight emotions—Neutral, Joy, Anger, Panic, Fear, Anxiety, Sadness, and Shame—across seven daily actions, including sitting, walking, and lifting. Data were recorded using the Xsens MVN system, capturing 28 3D joints at a frequency of 120 Hz.

\subsubsection{KDAE} The Kinematic Dataset of Actors Expressing Emotions \cite{zhang2020kinematic} was collected using a portable motion capture system that tracked 72 body markers at a frequency of 125 Hz. This dataset includes 1,402 recordings of seven emotions: Happiness, Sadness, Neutral, Anger, Disgust, Fear, and Surprise, performed by 22 semi-professional actors. For our analysis, we excluded hand joints and focused on 24 full-body joints.

\subsubsection{EGBM} The Emotional Gestures and Body Movements Corpus \cite{sapinski2019multimodal} contains 560 samples recorded by a Kinect V2 camera at 30 Hz. Sixteen Polish professional actors, evenly split between men and women aged 25 to 64, expressed seven emotions: Happiness, Sadness, Neutral, Anger, Disgust, Fear, and Surprise. Each emotion is represented by 80 samples, with 3D positions for 25 joints.

\subsubsection{Emotion Description} The aforementioned datasets only include emotional labels without detailed descriptions of emotional actions. To fully leverage the generative capabilities of LLMs, we manually annotated a subset of new data with fine-grained descriptions of emotional actions. For annotation, we referred to the emotional action descriptions provided in \cite{noroozi2018survey}. In total, 174 samples from the Emilya dataset and 105 samples from the KDAE dataset were labeled with these detailed descriptions.

\subsection{Implementation Details}

All experiments are conducted on NVIDIA 4$\times$A100 GPUs and are implemented using the PyTorch framework \cite{paszke2019pytorch}. We use LLaMA-7B \cite{touvron2023llama} as the base LLMs. For data preprocessing, we followed the scheme outlined in CTR-GCN \cite{chen2021channel}, adjusting each sample to 64 frames by padding shorter sequences with previous frames and downsampling longer sequences. 

\subsubsection{Training} For emotion description, the model is trained for 10,000 steps using labeled datasets. The LoRA parameters are configured with a rank of 64, an alpha of 16, and a dropout rate of 0.05. The global batch size is 16, and the maximum learning rate is 1e-5. For emotion recognition, the model undergoes 800,000 training steps, with the global batch size increased to 64. The LoRA parameters and maximum learning rate remain the same as in the emotion description stage. During skeleton encoder pre-training, we optimize the model for 200 epochs using the stochastic gradient descent (SGD) optimizer with an initial learning rate of 0.1 and a batch size of 64. The learning rate is reduced by a factor of 10 at epochs 100, 150, and 175. A warm-up strategy is applied for the first 5 epochs.

\subsubsection{Evaluation Protocols} We randomly split the dataset into training and testing sets at a 4:1 ratio. In emotion recognition, we extract the emotion labels from the generated sentences and compare them with the ground truth labels to compute accuracy. It is important to note that if a sentence contains multiple emotion labels or no emotion label, it will be marked as 'Error'. However, if the labels are synonyms or different grammatical forms of the same label, the recognition is deemed successful. For emotion description, we use three commonly employed metrics in natural language processing: Rouge, BLEU, and METEOR.

Further information regarding implementation details, novel LLMs, new skeleton encoder, and ablation study are provided in the supplementary material.

\begin{table}[tp]  
	\setlength{\tabcolsep}{2mm}
	\renewcommand\arraystretch{1.5}
	\centering  
	\fontsize{9}{9}\selectfont 
	\begin{threeparttable}  
		
		\begin{tabular}{@{}c|c|c|c@{}}
			\toprule
			Dataset    & Spatial Tokens & Temporal Tokens & Accuracy \\ \midrule
			\multirow{3}{*}{Emilya} & $\checkmark$ &                              & 85.08  \\ 
			&   & $\checkmark$  & 83.62 \\
			& $\checkmark$  & $\checkmark$ & \textbf{86.36}                     \\ \cmidrule(l){1-4}
			\multirow{3}{*}{KDAE}   & $\checkmark$ &                              & 62.99  \\
			& & $\checkmark$ & 61.21 \\
			& $\checkmark$ & $\checkmark$ & \textbf{63.35}             \\ \cmidrule(l){1-4}
			\multirow{3}{*}{EGBM}   & $\checkmark$  &                              & 53.21 \\
			&                            & $\checkmark$                          & 49.54                      \\
			& $\checkmark$                        & $\checkmark$                          & \textbf{55.96}                   \\ \bottomrule
		\end{tabular}
		
	\end{threeparttable}
	\caption{Classification results using spatial tokens and temporal tokens.}  
	\label{ST_token_result} 
\end{table}

\begin{table}[tp]  
	\setlength{\tabcolsep}{2mm}
	\renewcommand\arraystretch{1.5}
	\centering  
	\fontsize{9}{9}\selectfont 
	\begin{threeparttable}  
		
		\begin{tabular}{@{}c|c|c|c@{}}
			\toprule
			Token Type                      & Dataset& Separate & Joint\\ \midrule
			\multirow{3}{*}{Semantic}        & Emilya                      & 80.63                        & \textbf{85.44}              \\
			& KDAE                        & 67.97                        & \textbf{71.17}              \\
			& EGBM                        & 61.47                        & \textbf{66.97}              \\ \cmidrule(l){1-4}
			\multirow{3}{*}{Spatio-temporal} & Emilya                      & 86.36                        & \textbf{86.42}              \\
			& KDAE                        & \textbf{63.35}               & 62.28                       \\
			& EGBM                        & 55.96                        & \textbf{63.30}              \\ \bottomrule 
		\end{tabular}
		
	\end{threeparttable}
	\caption{Classification results for various training strategies.}  
	\label{Training_Strategies} 
\end{table}

\begin{table*}[tp]  
	\setlength{\tabcolsep}{2mm}
	\renewcommand\arraystretch{1.5}
	\centering  
	\fontsize{9}{9}\selectfont 
	\begin{threeparttable}  
		
		\begin{tabular}{c|c|c|ccc|c|ccc}
			\toprule
			\multicolumn{1}{c}{} &       & \multicolumn{4}{c|}{Emotion Description Task} & \multicolumn{4}{c}{Emotion Recognition Task} \\
			\midrule
			\multicolumn{1}{c|}{Token Type} & \multicolumn{1}{l|}{Order} & \multicolumn{1}{l|}{Accuracy} & \multicolumn{1}{l}{Rouge} & \multicolumn{1}{l}{BLEU} & \multicolumn{1}{l|}{METEOR} & \multicolumn{1}{l|}{Accuracy} & \multicolumn{1}{l}{Rouge-1} & \multicolumn{1}{l}{BLEU} & \multicolumn{1}{l}{METEOR} \\
			\midrule
			Semantic & D$\rightarrow$R   & N/A   & 0.1939  & 0.1301  & 0.2632  & 81.30  & 0.2018  & 0.1177  & 0.2148  \\
			Semantic & R$\rightarrow$D   & 82.04 & \multicolumn{3}{c|}{N/A} & 44.98 & 0.3138  & 0.2488  & 0.3408  \\ \cmidrule(l){1-10}
			Spatio-temporal & D$\rightarrow$R   & N/A   & 0.2774  & 0.2239  & 0.2941  & 81.40 & 0.2065  & 0.0401  & 0.0735  \\
			Spatio-temporal & R$\rightarrow$D   & 81.45 & \multicolumn{3}{c|}{N/A} & 19.59 & 0.2714  & 0.2149  & 0.2849  \\
			\bottomrule
		\end{tabular}%
	\end{threeparttable}
	\caption{Comparison of classification and emotion description results under different training orders. D$\rightarrow$R indicates that the emotion description task is completed first, followed by the recognition task, while R$\rightarrow$D indicates the reverse order. }  
	\label{G-R} 
\end{table*}

\subsection{Comparisons for Emotion Recognition}\label{ER}

\subsubsection{Evaluation on Spatio-temporal Tokens} We evaluate the classification results utilizing spatial and temporal tokens both separately and in combination, as presented in Tab. \ref{ST_token_result}. The results indicate that the separate classification of spatial and temporal tokens leads to lower accuracy when compared to their combined application across all three datasets. The integration of spatial and temporal tokens provides the model with more comprehensive information, thereby validating the effectiveness of our proposed multi-granularity skeleton tokenizer.

\subsubsection{Evaluation of Different Training Strategies} We evaluate the effectiveness of separate training strategy and joint training strategies with UST module by using skeleton tokens at different granularities. The results, presented in Tab. \ref{Training_Strategies}, show that using semantic tokens with joint training significantly improved classification accuracy across all three datasets, with an average increase of 4.5\%. Specifically, when employing spatio-temporal tokens, the accuracy on the EGBM dataset increased by 7.34\%. However, improvements are less pronounced for other datasets, with a slight decrease in accuracy observed on the KDAE dataset. These results suggest that joint training with semantic tokens enhances knowledge transfer between datasets, leading to better classification accuracy for individual datasets.

\begin{figure}[tp]
	\centering
	\begin{minipage}{0.49\linewidth}
		\centerline{\includegraphics[height=1.6in]{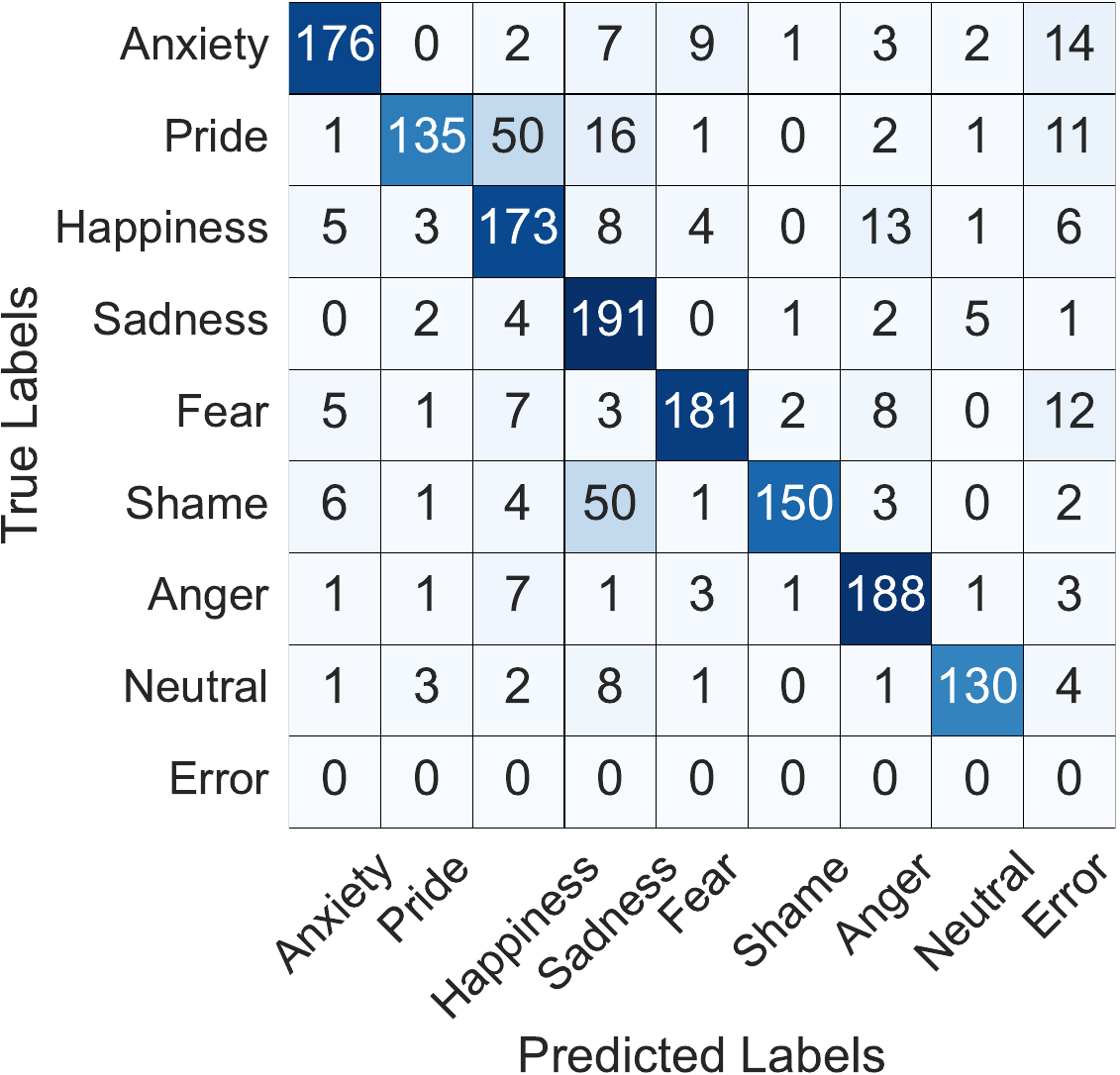}}
		\centerline{(a) Separate}
	\end{minipage}
	\begin{minipage}{0.49\linewidth}
		\centerline{\includegraphics[height=1.6in]{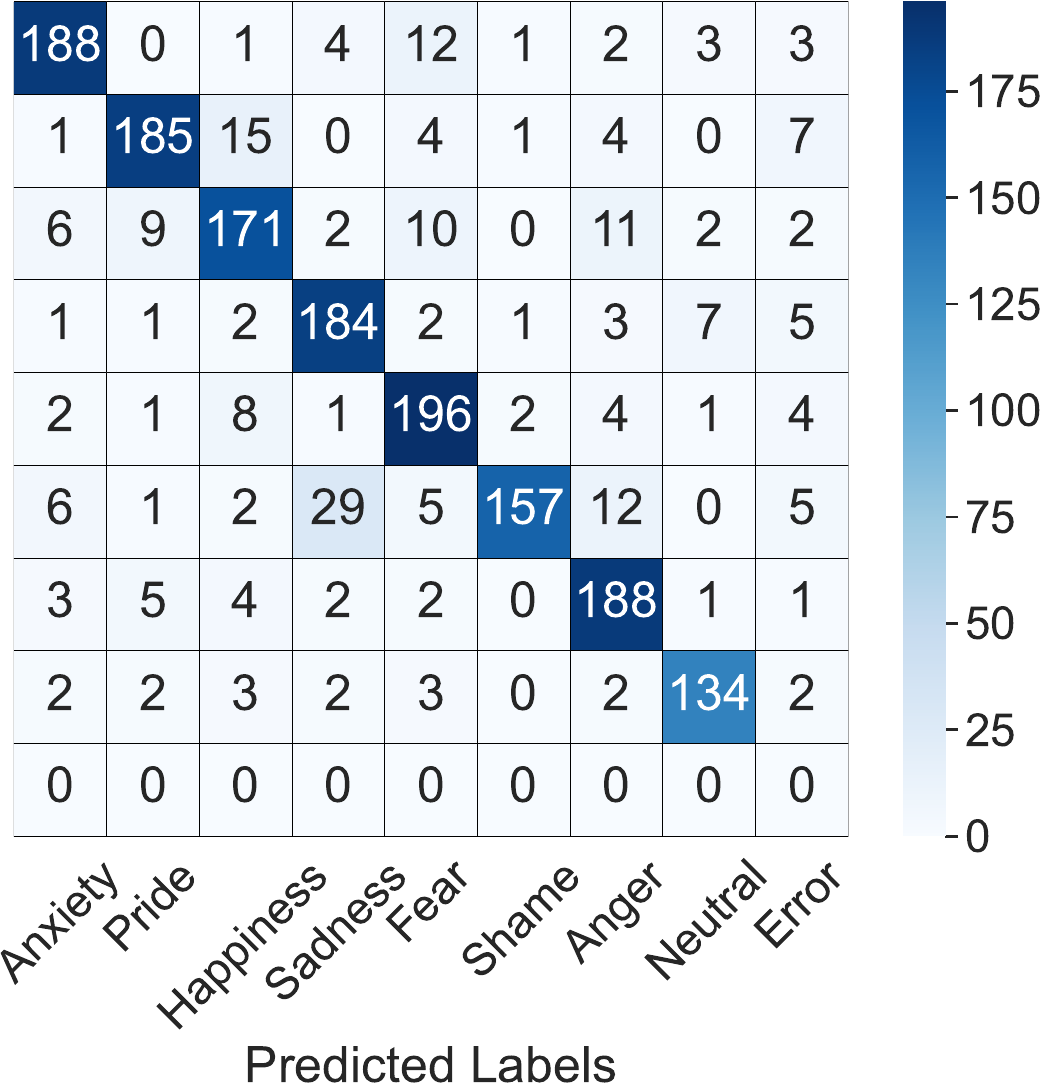}}
		\centerline{(b) Joint}
	\end{minipage}
	
	\caption{Confusion matrices for Emilya dataset using different training strategies.}
	\label{Qualitative_show}
\end{figure}

To further investigate the impact of UST module on recognition results, we plot the confusion matrix for the Emilya dataset, which has the largest number of samples. As shown in Fig. \ref{Qualitative_show}, the classification accuracy for anxiety, shame, and pride has significantly improved after employing the joint training strategy. Notably, these three emotions are absent in the KDAE and EGBM datasets. Compared to separate training, the joint training strategy has enhanced the model's recognition of emotions common to all three datasets, thereby positively influencing the recognition of emotion labels unique to the Emilya dataset.

\subsection{Comparisons for Emotion Description}

In this section, we compare the impact of semantic tokens and spatio-temporal tokens on emotion description and recognition results. Our model is designed to handle both emotion description and emotion recognition tasks simultaneously, so we also report the effects of different training orders on the results. Note that all results are obtained under joint training strategy, and both accuracy and generation metrics are averaged across the three datasets. As shown in Tab. \ref{G-R}, we can draw three conclusions: 

\begin{itemize}
	
	\item \textbf{In the emotion description task, spatio-temporal tokens outperform semantic tokens, whereas in the emotion recognition task, the opposite is true.} This advantage is likely because spatio-temporal tokens retain richer spatial and temporal information, which enhances the detail and quality of the generated descriptions. Semantic tokens are more concise and contribute better to classification.
	
	\item \textbf{The training order of different tasks significantly affects the recognition results.} When training the emotion description and emotion recognition tasks separately, good results can be achieved regardless of which tokens are used. However, when the emotion recognition task is completed first, and then the emotion description task is fine-tuned, catastrophic forgetting occurs. This results in a significant drop in average recognition accuracy by 37.06\% and 61.86\% for semantic and spatio-temporal tokens, respectively. This issue may arise from conflicting demands within the model for both emotion recognition and description capabilities. EAI-LLM imposes stringent constraints on recognition outcomes, expecting results to be framed in specific sentences, such as "This is a happy person.". These constraints often conflict with the diverse requirements of generation tasks, leading to reduced recognition accuracy when generation tasks are performed after recognition tasks.
	
	\item \textbf{Semantic tokens offer a balanced trade-off between recognition accuracy and description capability.} Although semantic tokens show lower accuracy compare to spatio-temporal tokens in emotion recognition tasks, their performance in subsequent emotion description tasks does not decline significantly, thereby maintaining relatively good accuracy.
	
\end{itemize}

\begin{figure}[tp]
	\centering
	\centerline{\includegraphics[width=3.3in]{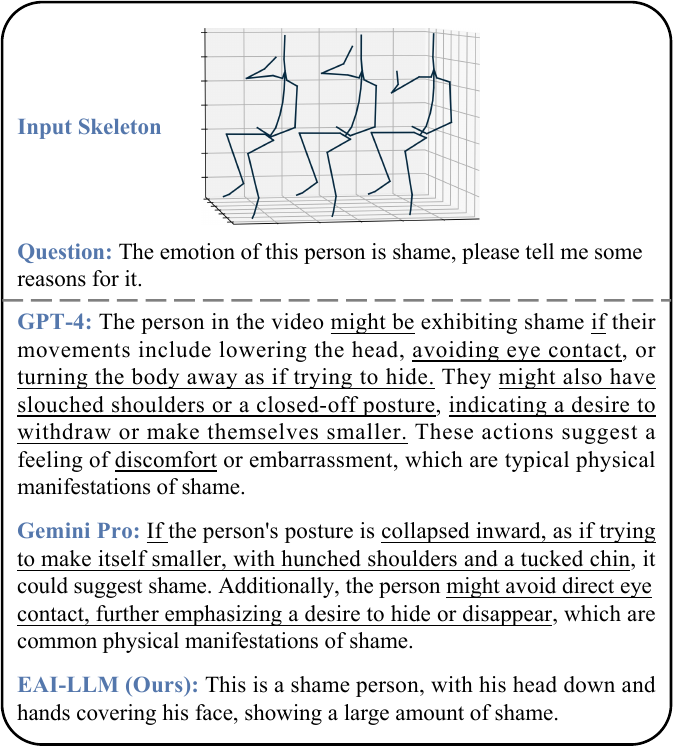}}
	\caption{Examples for emotion description capabilities of EAI-LLM. Underlined \underline{text} indicates that the description is unrelated to the input sequences.}
	\label{Gen_with_other_API}
\end{figure} 

\subsection{Comparison with the State-of-the-Art}

\begin{table}[tp]  
	\setlength{\tabcolsep}{2mm}
	\renewcommand\arraystretch{1.5}
	\centering  
	\fontsize{9}{9}\selectfont 
	\begin{threeparttable}  
		
		\begin{tabular}{@{}l|ccc|c@{}}
			\toprule
			& Emilya  & KDAE                                 & EGBM   &  D                              \\ \midrule
			AGCN \cite{2sagcn2019cvpr}   & 88.92                     & 56.58                              & 22.94        &  $\times$                 \\
			CTR-GCN \cite{chen2021channel} & \textbf{89.77}                     & \underline{70.46}                              & 63.30             &  $\times$                    \\
			GAP \cite{xiang2023generative} & \underline{89.16}                        & 67.26                              & \underline{66.06}                &  $\times$                   \\
			
			\textbf{EAI-LLM (Ours) }    & 85.44 & \textbf{71.17} & \textbf{66.97}  &  $\checkmark$    \\ \bottomrule
		\end{tabular}
	\end{threeparttable}
	\caption{Comparisons of emotion recognition capabilities with the state-of-the-art methods. Bold and underline indicate the best and the second best result. D represents emotion description ability from 3D skeleton sequences.}  
	\label{SOTA_R} 
\end{table}

Due to the limited reporting on skeleton-based emotion recognition and the absence of a standardized evaluation method, we have chosen to compare our approach with existing methods used in skeleton-based action recognition, which are analogous to emotion recognition. To ensure a fair comparison, we re-implemented all methods and standardized the data preprocessing procedures. As shown in Tab. \ref{SOTA_R}, our method demonstrates accuracy comparable to existing approaches on the KDAE and EGBM datasets, with the best accuracy reaching 71.17\% and 66.97\%, respectively. However, on the Emilya dataset, the accuracy only reached 85.44\%, still lagging behind other methods.

During the annotation process of the emotion description dataset, we converte the skeleton sequences into videos and then labeled the emotional action features by viewing these videos. Detailed examples are provided in the supplementary materials. Since current multi-modal large models do not support direct input of the original skeleton sequences, we instead input the videos derived from these skeleton sequences into the multi-modal LLMs, using the same prompts as EAI-LLM. As shown in Tab. \ref{SOTA_G}, when using the same prompt words, the metrics produced by our method surpass those generated by popular multi-modal LLMs.
\begin{table}[tp]  
	\setlength{\tabcolsep}{2mm}
	\renewcommand\arraystretch{1.5}
	\centering  
	\fontsize{9}{9}\selectfont 
	\begin{threeparttable}  
		
		\begin{tabular}{@{}c|ccc|c@{}}
			\toprule
			& Rouge           & BLEU            & METEOR     & R     \\ \midrule
			GPT-4       & 0.1282          & 0.0686          & 0.2006  &  $\times$        \\
			Gemini 1.5 Pro  & 0.1007          & 0.0596          & 0.1823  &  $\times$        \\
			\textbf{EAI-LLM (Ours) }          & \textbf{0.2018} & \textbf{0.1177} & \textbf{0.2148} &  $\checkmark$ \\ \bottomrule
		\end{tabular}                        
	\end{threeparttable}
	\caption{Comparison of emotion description capabilities with mainstream multimodal LLMs. R represents emotion recognition ability from 3D skeleton sequences.}  
	\label{SOTA_G} 
\end{table}

We present the visualization results in Fig. \ref{Gen_with_other_API}. The emotion descriptions generated by EAI-LLM exhibit significant consistency with the input sequences. In contrast, while both GPT-4 \cite{GPT4} and Gemini 1.5 Pro \cite{gemini15pro2024} can also generate emotion descriptions from video, these descriptions largely reflect the models' background knowledge and do not align as well with the input sequences.

\subsection{Limitations}

Since EAI-LLM is built upon LLMs, it inherits the LLM's limitations, such as hallucination, which can lead to ambiguous outputs. For instance, when given an input labeled as "Anxiety", EAI-LLM might generate: 

\begin{itemize}
	
	\item "This person is experiencing anguish or distress, as denoted by the frown and furrowed brow." While this sentence describes the state of anxiety, it does not explicitly mention the word "anxiety," and therefore fails to meet our accuracy criteria.
	
	\item "This person is expressing anxiety or fear." The output contains multiple emotional labels and cannot be clearly categorized under a specific emotional label, thereby resulting in a recognition failure.
	
\end{itemize}

Hallucinations in skeleton-based fine-grained emotion descriptions remain an unresolved issue. Employing a new evaluation protocol with a semantic detection module may be a potential solution.

\section{Conclusion}\label{Conclusion}

In this paper, we have introduced a novel emotion recognition model, EAI-LLM, designed to identify emotions from 3D full-body skeleton sequences and generate detailed emotion descriptions. We have used LLaMA as the language decoder and fine-tuned it on prompt-based question-answering tasks. Specifically, we have extracted skeleton tokens at various granularities to enable our model to produce more nuanced emotion descriptions. Additionally, the unified skeleton token module has significantly enhanced the accuracy of emotion recognition on individual datasets. Experimental results have demonstrated that our model effectively recognizes emotions from skeleton data and generates detailed emotion descriptions, even with limited data annotations.

\section*{Acknowledgments}
This work was supported in part by the National Natural Science Foundation of China (NSFC) under Grant 62072190 and 62202311, and in part by the Shenzhen Natural Science Foundation (the Stable Support Plan Program) under Grant 20220809180405001, and in part by the Innovation Team Project of Guangdong Province of China 2024KCXTD017, and in part by the Excellent Science and Technology Creative Talent Training Program of Shenzhen Municipality under Grant RCBS20221008093224017, and in part by the Guangdong Basic and Applied Basic Research Foundation under Grants 2023A1515011512, and in part by the Key Scientific Research Project of the Department of Education of Guangdong Province 2024ZDZX3012.

\bibliography{aaai25}

\section*{Appendices}

In this appendix, we present an ablation study on our model architecture (Appendix A) and examine the impact of different skeleton encoder settings and various large language models (LLMs) for the emotion recognition task in Appendices B and C, respectively. Additionally, we compare the effectiveness of different output formats for emotion recognition in Appendix D. Unless otherwise specified, and all experimental results are obtained using semantic tokens on the Emilya dataset.

\subsection*{A. Ablation on Architecture Designs}

Our model consists of three main components: the skeleton encoder, the linear layer, and the LLMs. The skeleton encoder is responsible for converting raw skeleton data into skeleton features. The linear layer then transforms these features into tokens that are more easily processed by the LLMs, which perform the emotion recognition and description task. We conduct ablation studies to assess the effectiveness of each module. Three key conclusions can be drawn from Tab. \ref{ablation_study_arch}.

\begin{itemize}
	
	\item Freezing the skeleton encoder's parameters is crucial. When the skeleton encoder's parameters are updated through fine-tuning, our model struggles to learn effective features, and the accuracy of the emotion recognition task does not exceed 14\%. Howlever, freezing the encoder's parameters leads to a significant improvement in the performance of the emotion recognition task.
	
	\item Fine-tuning the LLMs is necessary. After fine-tuning, the accuracy of the emotion recognition task increased from 73.75\% to 80.63\%.
	
	\item Increasing the number of linear layers is not always beneficial. As the number of linear layers increases, the accuracy of the emotion recognition task decreases from 80.63\% to 78.26\%. This suggests that a single projection layer is sufficient to align the skeleton encoder with the LLMs. 
	
\end{itemize}

Finally, we choose to freeze the skeleton encoder's parameters, use a single linear layer, and apply the LoRA to fine-tune the LLMs in our experiments.

\begin{table}[tp]  
	\setlength{\tabcolsep}{3mm}
	\renewcommand\arraystretch{1.5}
	\centering  
	\fontsize{9}{9}\selectfont 
	\begin{threeparttable}  
		\begin{tabular}{@{}l|c|c|c@{}}
			\toprule
			Skeleton Encoder   & Linear layer & LLMs   & Accuracy \\ \midrule
			Finetune & 1       & Freeze & 12.67    \\
			Finetune & 1       & LoRA   & 13.82    \\
			Freeze   & 1       & Freeze & 73.75   \\
			Freeze   & 1       & LoRA   & \textbf{80.63}   \\
			Freeze   & 2      & LoRA   & 80.39   \\
			Freeze   & 3      & LoRA   & 78.26   \\ \bottomrule
		\end{tabular}	
	\end{threeparttable}  
	\caption{Ablation on architecture designs.}
	\label{ablation_study_arch} 
\end{table}

\subsection*{B. Effectiveness of Skeleton Encoder Setting}

In Section 3.4, we use the semantic features to guide the skeleton encoding process, producing tokens that are more easily recognized by the LLMs. Here, we elaborate on the impact of semantic guidance on the emotion recognition task. As shown in Tab. \ref{ablation_study_loss}, we draw the following conclusions:

\begin{itemize}
	
	\item When using the raw skeleton encoder to obtain tokens, our model's classification accuracy is only 46.29\%. This suggests that the features extracted by the raw encoder are not well-aligned with the language space, making them difficult for the LLMs to recognize.
	
	\item Even without the cross-entropy guidance, tokens extracted solely through semantic guidance demonstrate some recognition capability. Specifically, optimized semantic tokens  achieve an accuracy of 78.81\%, and spatio-temporal tokens reach an accuracy of 76.43\%. This indicates that semantic guidance helps align the skeleton features closer to the language space, but these features alone lack sufficient differentiation for effective classification.
	
	\item The cross-entropy guidance proves to be crucial for enhancing the accuracy of the emotion recognition task. When the cross-entropy and semantic alignment are used together, the accuracy of the emotion recognition task improves further. This combination enables the extracted skeleton tokens to reside within the semantic space while also possessing a certain classification capability.
	
\end{itemize}

\begin{table}[tp]  
	\setlength{\tabcolsep}{5mm}
	\renewcommand\arraystretch{1.5}
	\centering  
	\fontsize{9}{9}\selectfont 
	\begin{threeparttable}  
		\begin{tabular}{@{}c|c|c|c@{}}
			\toprule
			$\mathcal{L}_{ce}$ & $\mathcal{L}_{se}$ & $\mathcal{L}_{st}$ & Accuracy \\ \midrule
			\checkmark &     &     & 46.29    \\ 
			& \checkmark   &     & 78.81   \\
			&     & \checkmark   & 76.43    \\
			\checkmark & \checkmark   &     & 80.63   \\
			\checkmark &     & \checkmark   & \textbf{86.36}   \\
			\checkmark &  \checkmark   & \checkmark   & 85.57   \\ \bottomrule
		\end{tabular}
	\end{threeparttable} 
	\caption{Performance comparison of different pre-training loss functions.} 
	\label{ablation_study_loss}
\end{table}

We investigate the impact of different skeleton encoders on our results by selecting three classic models in skeleton data processing as skeleton encoders: AGCN \cite{2sagcn2019cvpr}, CTR-GCN \cite{chen2021channel}, and HD-GCN \cite{Lee_2023_ICCV}. Tab. \ref{ablation_study_sk} presents the results of separate training and joint training using UST module. As shown in Tab. \ref{ablation_study_sk}, the accuracy of the emotion recognition task on the Emilya dataset exceeds 80\% when a separate training strategy is used. However, due to the limited sample sizes of the KDAE and EGBM datasets, the skeleton encoders are not fully trained, resulting in features that are not well-recognized by the LLMs. In the joint training strategy, CTR-GCN encoder performs well across all three datasets, with its accuracy on the Emilya dataset being only 0.07\% lower than that of HD-GCN encoder. Overall, CTR-GCN encoder demonstrates relatively stable performance and is therefore selected as the primary skeleton encoder for this study.
\begin{table}[tp]  
	\setlength{\tabcolsep}{5mm}
	\renewcommand\arraystretch{1.5}
	\centering  
	\fontsize{9}{9}\selectfont 
	\begin{threeparttable}  
		\begin{tabular}{@{}l|c|c|c@{}}
			\toprule
			Backbone    &  Dataset & Separate & Joint  \\ \midrule
			\multirow{3}{*}{CTR-GCN} & Emilya & 80.63  & 85.44 \\
			& KDAE   & 67.97  & 71.17 \\
			& EGBM   & 61.47  & 66.97 \\ \cmidrule(r){1-4}
			\multirow{3}{*}{AGCN}    & Emilya & 85.51  & 79.42 \\
			& KDAE   & 52.31  & 56.58 \\
			& EGBM   & 19.27  & 22.94 \\ \cmidrule(r){1-4}
			\multirow{3}{*}{HD-GCN}  & Emilya & 83.37  & 85.51 \\
			& KDAE   & 54.80  & 64.77 \\
			& EGBM   & 52.29  & 67.89 \\  \bottomrule 
		\end{tabular}
	\end{threeparttable}  
	\caption{Performance comparison of different skeleton encoder backbone.}
	\label{ablation_study_sk} 
\end{table}

\subsection*{C. Effectiveness of LLMs Setting}\label{conset}

Previous studies have demonstrated that LLMs can provide prior knowledge about body language \cite{qiu2024language, li2023large}. In this context, we experiment with different LLMs, including LLaMA-7B, LLaMA-13B \cite{touvron2023llama}, and Vicuna \cite{vicuna2023}, to evaluate their impact on emotion recognition. As shown in Tab. \ref{ablation_LLMs}, when using the same LoRA parameters ($r_{LoRA}$ and $\alpha_{LoRA}$ set to 64 and 16, respectively), the recognition accuracy of LLaMA-13B decreases by 4.81\% compared to LLaMA-7B. Only after adjusting to larger parameters does LLaMA-13B's accuracy surpass that of LLaMA-7B. However, due to the greater resources and additional time required to train the LLaMA-13B model, and given its marginal performance improvement, we choose to use LLaMA-7B as the base model in our subsequent experiments.

\begin{table}[tp]  
	\setlength{\tabcolsep}{5mm}
	\renewcommand\arraystretch{1.5}
	\centering  
	\fontsize{9}{9}\selectfont 
	\begin{threeparttable}  
		\begin{tabular}{@{}l|c|c|c@{}}
			\toprule
			Model    & $r_{LoRA}$ & $\alpha_{LoRA}$ & Accuracy \\ \midrule
			LLAMA 7B  & 64      & 16    & 80.63  \\
			Vicuna 7B & 64      & 16    & 74.85  \\
			LLAMA 13B & 64      & 16    & 75.82  \\
			LLAMA 13B & 128     & 32    & 81.06  \\ \bottomrule
		\end{tabular}	
	\end{threeparttable}  
	\caption{Performance comparison of different LLMs.}
	\label{ablation_LLMs} 
\end{table}

\subsection*{D. Effectiveness of Output Setting}\label{visual}

In emotion recognition task, it is essential for the model to generate sentences in a fixed format from which emotion labels can be extracted. To explore the impact of different output formats on the recognition task, we test three formats:
\begin{itemize}
	\item Output A: [Label]
	\item Output B: This is a/an [Label] person.
	\item Output C: This is a 3D skeleton sequence of a person. From their movements, it can be observed that their emotion is [Label].
\end{itemize}
In these formats, the emotion labels within the brackets are generated by the model. As shown in Tab. \ref{ablation_study_output}, when the model generates emotion labels directly or produces short sentences, the accuracy of the emotion recognition task is 80.94\% and 80.63\%, respectively, with no significant difference. However, when the model generates longer sentences, the accuracy of the emotion recognition task decreases by approximately 15\%. This decline can be attributed to the strict format requirements for sentence generation in recognition task. As the sentences become longer, the proportion of the sentence that constitutes the emotion label decreases, causing the model's attention to be dispersed, which in turn hinders the accurate identification of emotion labels.

\begin{table}[tp]  
	\setlength{\tabcolsep}{5mm}
	\renewcommand\arraystretch{1.5}
	\centering  
	\fontsize{9}{9}\selectfont 
	\begin{threeparttable}  
		\begin{tabular}{@{}cccc@{}}
			\toprule
			\multirow{1}{*}{} & \multicolumn{3}{c}{Output Format}      \\ \cmidrule(l){2-4} 
			& Output A     &Output B    &Output C      \\ \cmidrule(r){1-4}
			Accuracy            & \textbf{80.94}   & 80.63 & 65.23 \\ \bottomrule
		\end{tabular}
		
	\end{threeparttable}  
	\caption{Performance comparison of different output.}
	\label{ablation_study_output} 
\end{table}

\end{document}